\documentclass[%
reprint,
prl,
superscriptaddress,
longbibliography,
 amsmath,amssymb,
 aps,
]{revtex4-2}

\usepackage{graphicx}% Include figure files
\usepackage[caption=false]{subfig}
\usepackage{dcolumn}% Align table columns on decimal point
\usepackage{bm}% bold math
\usepackage{natbib}
\usepackage{changes}
\usepackage{color}
\usepackage{xcolor}
\usepackage{comment}

\newif\ifhighlightred
\highlightredfalse

\newcommand{\red}[1]{%
  \ifhighlightred{\color{red}#1}\else{#1}\fi%
}

\begin{document}

\preprint{APS/123-QED}

\title{High-Density Ultracold Neutron Source for Low-Energy Particle Physics Experiments}% Force line breaks with \\

\author{Skyler Degenkolb}
\email{Contact author: degenkolb@physi.uni-heidelberg.de}
\affiliation{Physikalisches Institut, Universität Heidelberg, Im Neuenheimer Feld 226, 69120 Heidelberg, Germany}
\affiliation{Institut Laue–Langevin, CS 20156, 38042 Grenoble Cedex 9, France}
\author{Estelle Chanel}
\email{Contact author: chanele@ill.fr}
\affiliation{Institut Laue–Langevin, CS 20156, 38042 Grenoble Cedex 9, France}
\author{Simon Baudoin}
\affiliation{Institut Laue–Langevin, CS 20156, 38042 Grenoble Cedex 9, France}
\author{Marie-H\'el\`ene Baurand}
\affiliation{Institut Laue–Langevin, CS 20156, 38042 Grenoble Cedex 9, France}
\author{Douglas H. Beck}
\affiliation{Department of Physics, University of Illinois at Urbana-Champaign, 1110 West Green Street, Urbana, IL 61801 USA}
\author{Juliette Bl\'e}
\affiliation{Institut Laue–Langevin, CS 20156, 38042 Grenoble Cedex 9, France}
\author{Eric Bourgeat-Lami}
\affiliation{Institut Laue–Langevin, CS 20156, 38042 Grenoble Cedex 9, France}
\author{Zeus Castillo}
\affiliation{Institut Laue–Langevin, CS 20156, 38042 Grenoble Cedex 9, France}
\author{Peter Fierlinger}
\affiliation{Physikdepartment, Technische Universit\"at M\"unchen, James-Franck-Straße 1, 85748 Garching bei München, Germany}
\author{Hanno Filter-Pieler}
\affiliation{Institut Laue–Langevin, CS 20156, 38042 Grenoble Cedex 9, France}
\author{Maurits van der Grinten}
\affiliation{Particle Physics Department, STFC Rutherford Appleton Laboratory, UK}
\author{Thomas Hepworth}
\affiliation{Physikalisches Institut, Universität Heidelberg, Im Neuenheimer Feld 226, 69120 Heidelberg, Germany}
\affiliation{Institut Laue–Langevin, CS 20156, 38042 Grenoble Cedex 9, France}
\author{Tobias Jenke}
\affiliation{Institut Laue–Langevin, CS 20156, 38042 Grenoble Cedex 9, France}
\author{Michael Jentschel}
\affiliation{Institut Laue–Langevin, CS 20156, 38042 Grenoble Cedex 9, France}
\author{Victorien Joyet}
\affiliation{Institut Laue–Langevin, CS 20156, 38042 Grenoble Cedex 9, France}
\author{Eddy Leli\`evre-Berna}
\affiliation{Institut Laue–Langevin, CS 20156, 38042 Grenoble Cedex 9, France}
\author{Husain Manasawala}
\affiliation{Physikalisches Institut, Universität Heidelberg, Im Neuenheimer Feld 226, 69120 Heidelberg, Germany}
\affiliation{Physikdepartment, Technische Universit\"at M\"unchen, James-Franck-Straße 1, 85748 Garching bei München, Germany}
\author{Thomas Neulinger}
\affiliation{FRM II, Lichtenbergstraße 1, 85748 Garching bei München, Germany}
\affiliation{Institut Laue–Langevin, CS 20156, 38042 Grenoble Cedex 9, France}
\author{Ulrich Schmidt}
\affiliation{Physikalisches Institut, Universität Heidelberg, Im Neuenheimer Feld 226, 69120 Heidelberg, Germany}
\author{Kseniia Svirina}
\affiliation{Institut Laue–Langevin, CS 20156, 38042 Grenoble Cedex 9, France}
\affiliation{Physikalisches Institut, Universität Heidelberg, Im Neuenheimer Feld 226, 69120 Heidelberg, Germany}
\author{Xavier Tonon}
\affiliation{Institut Laue–Langevin, CS 20156, 38042 Grenoble Cedex 9, France}
\author{Oliver Zimmer}
\affiliation{Institut Laue–Langevin, CS 20156, 38042 Grenoble Cedex 9, France}

\date{\today}% It is always \today, today,

\begin{abstract} 
SuperSUN, a new superthermal source of ultracold neutrons (UCN) at the Institut Laue-Langevin, exploits inelastic scattering of neutrons in isotopically pure superfluid $^4$He at temperatures below $0.6\,$K. For the first time, continuous operation with an intense broad-spectrum cold neutron beam is demonstrated over 60 days. We observe continuous UCN extraction rates of $21000\,$s$^{-1}$, and storage in the source with saturated density $273\,$cm$^{-3}$. The low-energy \textit{in-situ} UCN spectrum is alterable via accumulation and holding delays, opening new possibilities in fundamental and applied physics.
\end{abstract}

\maketitle

\textit{Introduction}\textemdash Ultracold neutrons (UCN) provide a route to high-precision experiments via long holding times \cite{steyerl,golub,ignatovich}.
Precision measurements and searches for new physics are thus possible with relatively small numbers of stored UCN, including measurements of the neutron's permanent electric dipole moment (EDM)~\cite{PSIEDM} and lifetime~\cite{PhysRevC.111.045501}, angular correlations in $\beta$ decay~\cite{UCNA}, bound states in Earth's gravity~\cite{Cronenberg:2018qxf,Jenke:2020obe}, tests of Lorentz invariance~\cite{PhysRevLett.103.081602,Ivanov:2020son}, searches for axion-like new particles ~\cite{Ayres:2023txi}, and limits on its oscillation to other neutral particles~\cite{PhysRevLett.131.191801}.
Nevertheless, low statistics has been a longstanding challenge of fundamental UCN science and a major obstacle to applications.
We present the full implementation of a 50-year-old concept to resolve this challenge \cite{GOLUB1977337}, following many proofs-of-principle \cite{AGERON1978469,golub1983operation,Brome:2001sm,BAKER200367,Zimmer:2007qw,PhysRevLett.107.134801,Masuda:2012tgd,Piegsa:2014kwa,TUCAN:2018vmr}.
Reconstructing stored UCN energy spectra, we show that experimentally quantified losses can be used to deliberately shape them.
This opens new possibilities to study and control spectrum-dependent systematics in UCN experiments.

\begin{figure}[h]
\includegraphics[clip,width=\columnwidth]{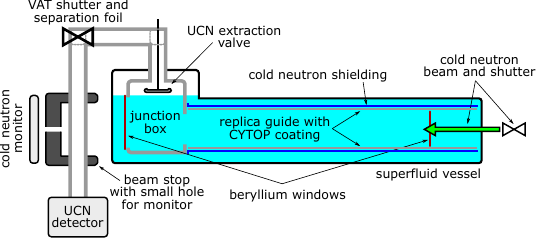}
\caption{\label{fig:apparatus} Diagram of SuperSUN, indicating its main neutron-handling elements. A 90$^\circ$ bend in the horizontal extraction guide, before the VAT shutter and out the page, displaces the vertical guide to the detector out of the cold neutron beam.}
\end{figure}

We distinguish two categories of UCN source: ``flux sources" delivering high particle current, and ``density sources" characterized by high numbers of stored neutrons per unit volume\textemdash especially \textit{in-situ}, i.e., within the source itself.
\red{High \textit{in-situ} densities build up over time, equilibrating high UCN production rate densities against low total loss rates.
Losses at material boundaries depend on UCN energy, through both the wall-interaction frequency and energy-dependent penetration into wall materials.}
Other losses, most importantly $\beta$ decay (loss rate $\Gamma_\beta \approx 0.0011\,$s$^{-1}$) and capture on nuclei in the bulk of storage volumes, are effectively energy-independent.
Stored UCN spectra thus evolve in time, and can vary greatly between sources and different operating conditions.
Technical choices such as materials and mechanical tolerances (gaps) strongly impact UCN loss.

Superfluid $^4$He is the only production medium within which UCN loss can, in principle, be arbitrarily reduced to approach the natural limit imposed by $\Gamma_\beta$.
The main UCN production channel is ``conversion" of a cold neutron with $8.9\,$\AA~wavelength, which transfers nearly its entire energy and momentum to the superfluid by creating a single phonon. 
The known UCN production cross-section and wavelength dependence \cite{BAKER200367,SCHMIDTWELLENBURG2009259} give an approximately isotropic distribution of wavevectors $\bm{\mathrm{k}}$ \cite{Lam95}, and a differential kinetic-energy spectrum scaling as $\sqrt{E_\text{k}}dE_\text{k} \propto|\bm{\mathrm{k}}|^2d|\bm{\mathrm{k}}|$ \cite{GOLUB1977337}.
Absorption losses are entirely absent in $^4$He, whereas $^3$He at a relative concentration $x_3$ induces a capture loss rate $\Gamma_3 = 2.4\times 10^7x_3\,$s$^{-1}$ \cite{nEDM:2019qgk}.
Up-scattering out of the UCN energy range is given by $\Gamma_T \approx 0.008\,T^7\,$s$^{-1}$ \cite{golub1979storage,Leung:2015gba}, with $\Gamma_T \ll \Gamma_\beta$ for $T\lesssim0.6\,$K.

We report initial measurements of superthermal UCN production from the SuperSUN source at the Institut Laue-Langevin (ILL) in Grenoble \cite{supersundata,PhysRevC.92.015501}.  After briefly describing the apparatus, \red{we quantify the main loss mechanisms affecting UCN accumulation and \textit{in-situ} storage within the source. We then reconstruct \textit{in-situ} total-energy spectra from the physical loss parameters obtained via fits to data, and show that very different spectra can be obtained as a consequence of different accumulation and holding delays.}
Total UCN production at saturation is 4.5 times that of the SUN-2 prototype source at the ILL \cite{panEDM,chanel2022concept}, exceeding earlier demonstrations by a factor 14 \cite{PhysRevLett.107.134801,Piegsa:2014kwa} and achieving the largest UCN density stored and measured to date.

\begin{figure}[tp]
\includegraphics[]{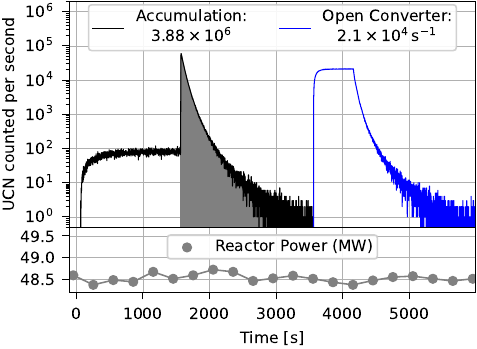}
\caption{\label{fig:modes} Example data in two operating modes. \textit{Left} (black): UCN accumulate for $1500$~s in the closed converter, with $\sim 80\,\text{s}^{-1}$ leaking through a monitoring hole in the UCN valve. The valve opens after the cold neutron beam is shut off: the number of extracted UCN is the shaded area under the peak, $3.88\times10^6$. \textit{Right} (blue): a lower, steady rate of $2.1\times10^4\,$s$^{-1}$ is continuously extracted via the open UCN valve, with beam on. After a chosen interval ($600\,$s), the beam is shut off.}
\end{figure}

\textit{Apparatus}\textemdash Figure~\ref{fig:apparatus} shows a diagram of SuperSUN.
 The cold neutron beamline H523 originates in a liquid deuterium moderator, and terminates at the UCN source \cite{chanel2022concept}.
 Gammas and short-wavelength neutrons are suppressed by a strongly-curved supermirror guide \cite{mezei1976novel,mezei1977corrigendum} with $m=1.2$ \footnote{The ‘‘$m$-value’’ of multilayer coatings used for neutron supermirrors is the factor by which the critical angle exceeds that of Ni with natural isotopic composition.}.
 A final guide section with $m=2.5$ and a tapered octagonal cross section~\cite{Deg18} adapts the rectangular guide to the circular cross-section of SuperSUN's converter, which is transversely bounded by a $3\,$m long replica-type guide \cite{plonka2007replika,serebrov2017replica} of \red{$R=37.2\,$mm inner radius}.
 %$74.4\,$mm inner diameter.
 The replica guide consists of a laser-cut, rolled, and laser-welded Ni foil $\sim 150\,${\textmu}m thick, with an $m=3$ supermirror coating on its inner surface consisting of about $600$ layers.
 Cold neutrons enter the converter through a 1~mm thick Be window, with a capture-weighted mean flux of $2.1\times10^{10}\,$cm$^{-2}$ s$^{-1}$ across all wavelengths \footnote{Measured at $56.9\,$MW reactor power. The $\sim0.1$ \AA~wavelength range for single-phonon UCN production near $8.9$ \AA~\cite{yoshiki2003cross} represents approximately $1\%$ of this total, with shorter wavelengths increasing the total UCN yield by $\sim10\%$ via multi-phonon processes \cite{SCHMIDTWELLENBURG2009259}.}.\nocite{yoshiki2003cross}
 Neutrons escaping through the replica foil are absorbed in 0.2~mm of metallic Gd wrapped around it.
 The converter volume is filled with superfluid $^4$He via a packed-alumina superleak \cite{Zimmer:2010zz,YOSHIKI2005399}, reducing $^3$He contamination.
To reduce UCN loss the cylindrical guide wall is coated with the fluoropolymer CYTOP~\cite{CYTOP}, with a neutron optical potential $V_\text{CYTOP}=115\,$neV and loss factor $f_\text{CYTOP} \leq (2.7\pm0.2)\times 10^{-5}$ at 10 K~\cite{Neu22}.
A Be-coated aluminum junction box couples this guide to a vertical extraction system.
Cold neutrons exit the converter through a second Be window.
A circular extraction aperture is closed during UCN accumulation and storage by a diamond-like carbon (DLC) coated aluminum disk.
A linear actuator lowers this disk valve to release stored UCN for experiments, with the free superfluid helium surface below its range of motion.
UCN leaving the helium can exit the converter through $50\,$mm diameter polished stainless steel guides, passing two 90$^\circ$ bends that incorporate Ge windows coated with deuterium-enriched DLC \footnote{Hydrogen impurities in DLC can significantly reduce its neutron optical potential. These DLC coatings are prepared with deuterated precursor chemicals, to maintain a high scattering length density despite impurities.}.
These windows reflect UCN while transmitting infrared thermal radiation out of the extraction guide, lowering the heat load on the converter.
We minimize mechanical gaps between neutron-optical components.

A recirculating $^3$He evaporation refrigerator absorbs heat delivered to the converter via mechanical supports, thermal radiation, and absorption of cold neutrons.
SuperSUN operates at $0.56\,$K with the beam shut off.
The cold neutron beam adds $20$-$25\,$~mW heating for nominal beam flux, increasing the converter temperature by about $30\,$mK and the (still negligible) loss rate $\Gamma_T$ by $40$\%. 
Thermal screens and internal liquid helium reservoirs are cooled by three cryocoolers (Sumitomo RDK-415D).
Liquid helium is filled from a dewar into a 100-liter internal reservoir, and cooled to superfluid by pumping a separate ``1~K pot".
A finned heat exchanger at the bottom of the $^3$He pumping column is used to cool purified $^4$He at the point where it arrives in the converter.
SuperSUN is brought into operation within 8 days and operates continuously for full reactor cycles, typically 7-9 weeks, with the liquid helium reservoir refilled weekly.

\begin{figure}[t]
\includegraphics[]{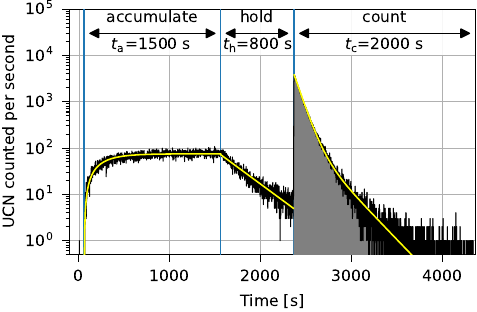}
\caption{\label{fig:delayed} Detected UCN rate during an accumulation-mode measurement, with extraction delayed by an $800$~s holding period after stopping UCN production.}
\end{figure}

\begin{figure}[tb]
\includegraphics[]{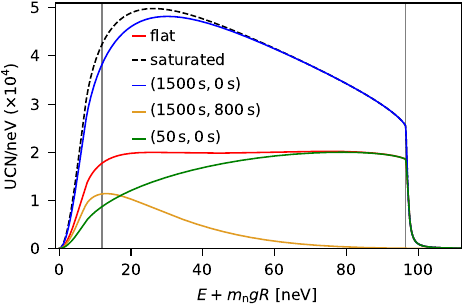}
\caption{\label{fig:spectra} UCN spectra for various choices of $(t_\text{a},t_\text{h})$. Nearly flat spectra can be obtained by performing two consecutive accumulate/hold sequences: $(1500\,\text{s},\, 800\,\text{s})$ and $(150\,\text{s},\, 0\,\text{s})$. The dashed curve gives the saturated maximum for each energy. Vertical gridlines show thresholds for extraction and storage.}
\end{figure}

\begin{figure}[th]
\includegraphics[]{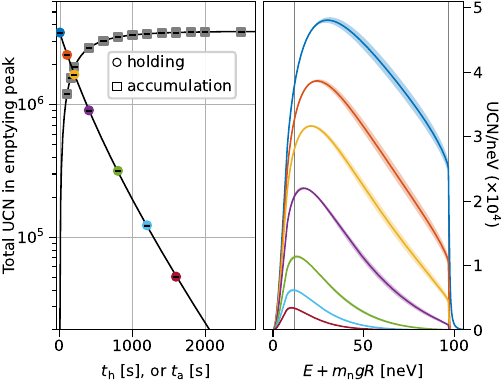}

\caption{\label{fig:storage} \textit{Left:} variation of total extracted UCN number with $t_\text{h}$ (holding: $t_\text{a}=1500\,$s fixed) or $t_\text{a}$ (accumulation: $t_\text{h}=0\,$s fixed), and curve fits to our physical model. Statistical error bars are shown within data markers. \textit{Right:} \textit{in-situ} total-energy spectra, reconstructed using Eq.~\eqref{eq:spectrum} and the fit values for $f$ and $\Gamma_\text{EI}$ (see text). The seven curves at right correspond, respectively, to the sequence of seven points in the holding scan at left. Colored bands show $1\sigma$ error ranges from the fit.}
\end{figure}

\textit{Characterization}\textemdash SuperSUN operates in two modes: accumulation and open converter, as shown in Figs.~\ref{fig:modes} and \ref{fig:delayed}.
After a 60 s background measurement (background rates are in fact negligible) we opened the cold neutron beam, and UCN accumulated in the converter for $t_\text{a}=1500\,$s with the extraction valve closed.
The detected leakage rate approached $80\,$s$^{-1}$ at saturation, from a small monitoring hole drilled in the valve disk.
We then shut the beam off, and opened the extraction valve after a $t_\text{h}=1\,$s holding delay.
UCN then exited the source, gaining $V_\text{LHe}=18.5\,$neV in kinetic energy by leaving the helium.
The bottom of a $50\,$mm diameter horizontal extraction guide is $26\,$cm above the replica guide axis.
We set zero potential to give a total energy $E=E_\mathrm{k}+m_\text{n}gz + \xi  V_\text{LHe}$, where $\xi=1$ in vacuum and $\xi=0$ in helium.
Gravity acts with $m_\text{n}g=102.6\,$neV/m, where $z$ is the height above the replica guide axis.
The minimum total energy needed to exit the converter is thus $E_\text{min}= m_n g \cdot 26\,\text{cm} -V_\text{LHe} = 8.1\,$neV, shown by vertical gridlines at $E_\text{min} + m_n g R = 12\,$neV in Figs.~\ref{fig:spectra} and \ref{fig:storage}.

After exiting the source UCN fell $82\,$cm in a vertical guide, exceeding the normal momentum needed to enter a DUNya-type $^3$He-based UCN detector \cite{groshev_1974} through a $100\,${\textmu}m aluminum window.
The total number of UCN detected in the $2000$~s counting interval was $3.88\times 10^6$ without any corrections, within 5\% of the design goal~\cite{panEDM}.
In open converter measurements, UCN are produced and extracted continuously.
The steady detected rate, $2.1\times10^4$~s$^{-1}$, is more than $2\times$ larger than the sum of all closed-converter losses (as extracted from fits, see below).
Extraction thus dominates loss from the open converter.

Around $1575\,$s in Fig.~\ref{fig:modes} the peak event rate $6.3\times 10^4$~s$^{-1}$ includes pileup, where a single detection event represents more than one neutron.
We apply pileup corrections, up to event multiplicity four and totaling less than $10\%$ for the highest rates, for fits to data and throughout Figs.~\ref{fig:delayed}-\ref{fig:storage}.
Corrections \textit{are not applied} to our values for total UCN output or density, which are therefore conservative.
The converter's $14.2$-liter volume gives an \textit{in-situ} saturated UCN density of 273 cm$^{-3}$.
To our knowledge this is the highest UCN density stored in any measurement to date.

Fig.~\ref{fig:delayed} shows UCN extraction after a longer holding delay ($t_\text{h}=800\,$s) without further accumulation, and Fig.~\ref{fig:spectra} shows calculated UCN spectra for different accumulation and holding sequences.
It is instructive to plot total extracted UCN number as a function of $t_\text{h}$ or $t_\text{a}$, on a common abscissa, as in Fig.~\ref{fig:storage}.
In each case the ordinate gives an integral of the entire counting peak. 
To compensate drifts of cold neutron beam intensity, each data point is scaled by the ratio of the average beam rate during accumulation to the average beam rate of the entire data series.
This correction is at most a few percent and typically much less, but also increases statistical errors.

UCN do not thermalize, but can approach ``mechanical equilibrium" during storage \cite{golub}, meaning that the accessible position-momentum phase space becomes uniformly occupied.
Due to gravity, spatially homogeneous UCN ensembles are \textit{not} mechanically equilibrated.
Randomizing interactions such as non-specular wall reflections favor equilibrium, but compete with loss rates in the sense that losses can outpace equilibration.
At equilibrium the time constant $\Gamma^{-1}$ for UCN loss is time-independent \cite{Pendlebury:1994he}. 
It does depend on $E$, which is effectively conserved, with inelastic interactions appearing as loss channels \footnote{We neglect microphonic heating, which could in principle lead to small energy shifts within the trapped spectrum, see, e.g., \cite{UCNTau:2018atb}.}.

UCN accumulation and storage are then described by exponential functions of the total loss rate $\Gamma(E;f,\Gamma_\text{EI})$, where $f$ and $\Gamma_\text{EI}$ are the free parameters for fitting to data.
Energy-dependent loss arises from a complex neutron optical potential $U=V\cdot(1-if)$, where $f$ is the dimensionless loss factor associated to a real potential $V$.
The sum of energy-independent losses is $\Gamma_\text{EI}\approx\Gamma_\beta + \Gamma_3$.
The differential total-energy spectrum is then
%i.e., the number $dN$ in a small interval $dE$, is
\begin{align}
\frac{dN}{dE} = \frac{P(E)e^{-\Gamma(E;f,\Gamma_\text{EI}) t_\text{h}}}{\Gamma(E;f,\Gamma_\text{EI})}  \left( 1 - e^{-\Gamma(E;f,\Gamma_\text{EI}) t_\text{a}} \right),
\label{eq:spectrum}
\end{align}
where $P(E)\propto\sqrt{E+m_\text{n}gR}\, \text{Re}\!\left[ _2F_1\!\left( -\frac{1}{2},\frac{3}{2},3, \frac{2m_\text{n}gR}{E+m_\text{n}gR}\right)\right]$ is the energy-dependent production rate 
and $_2F_1$ is the ordinary hypergeometric function.
Details of the model and fitting are discussed in Supplemental Material \cite{supp}, and our calculation of $\Gamma$ follows 
Ref.~\cite{Pendlebury:1994he}.
As Fig.~\ref{fig:spectra} shows, choice of $(t_\text{a},t_\text{h})$ significantly shapes the stored spectrum.

We fit the integral of Eq.~\eqref{eq:spectrum} %shifted by $V_\text{LHe}$,
to storage and accumulation data with various $(t_\text{a},t_\text{h})$, as shown in Fig.~\ref{fig:storage}.
The lower integration limit is $E_\text{min}$, while the upper limit $E_\text{max} = 1.5\cdot (V_\text{CYTOP}-V_\text{LHe})$ includes ``overcritical" UCN that can be transiently trapped.
Our fits include gravitational and geometrical effects, but simplify the converter geometry to a perfect cylinder.
We constrain $f$ to have the same value on the converter wall and end caps: any difference cannot be resolved from our data.
The ratio $\epsilon\approx 0.413$, of the known real potentials for CYTOP and Be, is a fixed parameter\textemdash as are the converter dimensions, and strength of gravity \cite{supp}.
Extraction and detection efficiencies are neglected, but can be reliably simulated \cite{Degenkolb:2025vql}.
For highly specular surfaces, simulations also indicate evolution towards anisotropic (non-equilibrated) ensembles with long-lived trajectories, characterized by low-angle or infrequent wall interactions.

The fit gives $f=(5.22\pm0.25)\times10^{-5}$ and $\Gamma_\text{EI}=(1.15\pm0.04)\times10^{-3}\,\text{s}^{-1}$.
While in phenomenological (e.g., bi-exponential) models, fits are biased by the distribution of measured $(t_\text{a},t_\text{h})$, we observe substantial consistency across different datasets.
%
%Differences can be explained by physically plausible changes of $f$ or $\Gamma_\text{EI}$, e.g., from small variations of $^{3}$He concentration.
%
Differences do not much exceed fit errors\textemdash for example, storage data later in the same reactor cycle give $f=(4.80\pm0.21)\times10^{-5}$ and $\Gamma_\text{EI}=(1.25\pm0.03)\times10^{-3}\,\text{s}^{-1}$.
While this might suggest slightly different $^{3}$He concentrations, structure in fit residuals ($\chi^2_\text{r}=2.7$ and $5.0$, respectively) also hints at limitations from assuming mechanical equilibrium \cite{supp}.

We use the same function to fit both accumulation and holding data, and to perform joint fits in which both $t_\text{a}$ and $t_\text{h}$ vary.
Given values of $f$ or $\Gamma_\text{EI}$, Eq.~\eqref{eq:spectrum} enables calculating the energy-dependent evolution of UCN spectra throughout accumulation and storage, as in Fig.~\ref{fig:spectra}.
In particular, we use $f$ and $\Gamma_\text{EI}$ from fits to reconstruct the \textit{in-situ} spectra from integrated counting data, for arbitrary ($t_\text{a},t_\text{h}$).
Fig.~\ref{fig:storage} shows this reconstruction at right, for the actually-measured holding times $t_\text{h}$ of the scan.

Given a value for the neutron lifetime, $\Gamma_\text{EI}$ can be used to set a limit on $^3$He concentration in the converter.
Conservatively taking $\Gamma_\beta=(895\,\text{s})^{-1}$, we obtain an upper limit of $x_3<3\times10^{-12}$ from the holding data of Fig.~\ref{fig:storage}.
For all datasets so far, $x_3\lesssim2\times10^{-11}$.
For $f$ we emphasize that losses through gaps, including the extraction valve hole, are absorbed additively in this effective parameter.
It therefore represents an upper bound on the loss factor of the actual wall materials.
A simple model where neutrons encountering any gap are lost with unit probability implies $f\geq5\times10^{-5}$, for $30\,${\textmu}m gaps at every mechanical junction.
Thus, fitted values of $f$ can also be interpreted as constraining the maximum gap size.

Our measurements were intentionally performed without a separation foil.
We interspersed repeated monitoring measurements, which show compounding losses of $\sim 3\%$ per day in total UCN output, likely due to freezing-in of gas impurities from the guides outside SuperSUN.
Based on demonstrations already performed at SUN-2 \cite{panEDM,chanel2022concept}, a $0.5\,${\textmu}m polypropylene foil is expected to reduce degradation to negligible levels at the expense of $10\%$ (static) reduction in total UCN output.

\textit{Discussion}\textemdash UCN produced in $^4$He have very low energy \cite{neulinger2024vertical}, and high phase-space density.
Apart from SuperSUN, all full-scale UCN sources to date are flux sources with harder UCN spectra \cite{STEYERL1986347,bison2023time,WONG2023168105,Abe:2026mwo} and more dilute phase space.
Our spectra are highly advantageous in storage experiments, complementing shorter and flow-through type measurements at flux sources.
Extracting UCN also enables external storage experiments, with significant gains balancing polarization and transport losses.
This approach is being developed at SuperSUN for PanEDM \cite{panEDM}, a neutron EDM measurement where $>250\,$s measurement duration and a $400\,$s repetition period will enable accumulating $75\%$ of the saturated maximum UCN number in SuperSUN.
Sensitivities projected from this performance can be globally leading \cite{panEDM}, and other storage experiments should profit similarly.

The source's full phase-space density, available for \textit{in-situ} measurements, is not presently reachable from any other UCN source or production technology.
Our analysis of extracted, integral UCN data also uniquely enables reconstructing \textit{in-situ} energy spectra, as well as producing and extracting novel (e.g., flat) spectra when the physical loss parameters are known.
This is key for: \textit{in-situ} measurements with significant potential gains \cite{GOLUB19941,nEDM:2019qgk,degenkolb2022approaches}, novel concepts like storage of gravitational quantum states \cite{Abele:2009dw} for fundamental physics and applications, and systematic studies involving UCN spectra.

\textit{Summary and outlook}\textemdash We reported first UCN production from the new source SuperSUN at the ILL.
The maximum number of UCN stored \textit{in-situ} was $3.88\times 10^6$, giving a highest-ever uncorrected density of 273 cm$^{-3}$.
We also showed how specific \textit{in-situ} storage measurements enable a detailed understanding UCN loss, and spectral evolution, in the converter.
The first use of SuperSUN will be supplying UCN to the PanEDM neutron electric dipole moment experiment.
A second phase will incorporate a superconducting octupole magnet, to generate a radial magnetic gradient with the field reaching 2.1~T at the boundary surface of the converter.
This gives a deeper trap for low-field seeking UCN, and increased loss for high-field seekers \cite{PhysRevC.92.015501}.
Polarized UCN will therefore accumulate in the converter, and the extraction system and transport guides will be modified to preserve UCN polarization.

\textit{Acknowledgements}\textemdash This project was funded by the Agence Nationale de la Recherche (ANR) under the grant ANR-14-CE33-0023-01, and by the Institut Laue–Langevin's Endurance program. We thank F. Waldherr, L. Dimmler, and S. Plantier for support and discussions, as well as the entire PanEDM collaboration, and technical services of the ILL, TUM, and Heidelberg University. We also gratefully acknowledge S-DH GmbH, for extensive joint work developing novel cold neutron guides for SuperSUN. TH is partially funded by the Heidelberg IMPRS \textit{Precision Tests of Fundamental Symmetries}.

\bibliography{bibliography}

@misc{supp,
note = {See Supplemental Material at [URL will be inserted by publisher] for details of the physical model and data fitting, as well as Ref. \cite{kennard} therein}
}

@article{UCNTau:2018atb,
    author = "Callahan, Nathan and others",
    collaboration = "UCNTau",
    title = "{Monte Carlo simulations of trapped ultracold neutrons in the UCN{\ensuremath{\tau}} experiment}",
    eprint = "1810.07691",
    archivePrefix = "arXiv",
    primaryClass = "physics.ins-det",
    doi = "10.1103/PhysRevC.100.015501",
    journal = "Phys. Rev. C",
    volume = "100",
    number = "1",
    pages = "015501",
    year = "2019"
}

@article{Abele:2009dw,
    author = "Abele, H. and Jenke, T. and Leeb, H. and Schmiedmayer, J.",
    title = "{Ramsey's Method of Separated Oscillating Fields and its Application to Gravitationally Induced Quantum Phaseshifts}",
    eprint = "0907.5447",
    archivePrefix = "arXiv",
    primaryClass = "nucl-ex",
    doi = "10.1103/PhysRevD.81.065019",
    journal = "Phys. Rev. D",
    volume = "81",
    pages = "065019",
    year = "2010"
}

@article{Abe:2026mwo,
    author = "Abe, K. and others",
    title = "{A New High-Intensity Source for Ultracold Neutrons}",
    eprint = "2607.03033",
    archivePrefix = "arXiv",
    primaryClass = "physics.ins-det",
    month = "7",
    year = "2026",
    journal ={}
}

@article{yoshiki2003cross,
  title={The cross sections for one phonon emission and absorption by slow neutrons in superfluid liquid Helium},
  author={Yoshiki, H},
  journal={Computer Physics Communications},
  volume={151},
  number={2},
  pages={141--148},
  year={2003},
  publisher={Elsevier}
}

@article{mezei1976novel,
  title={Novel polarized neutron devices: supermirror and spin component amplifier},
  author={Mezei, F},
  journal={Communications on Physics (London)},
  volume={1},
  number={3},
  year={1976},
  pages={81–85}
}

@article{mezei1977corrigendum,
  title={Corrigendum and first experimental evidence on neutron supermirrors},
  author={Mezei, F and Dagleish, PA},
  journal={Communications on Physics (London)},
  volume={2},
  year={1977},
  pages={41–43}
}

@article{Pendlebury:1994he,
    author = "Pendlebury, J. M. and Richardson, D. J.",
    title = "{Effects of gravity on the storage of ultracold neutrons}",
    doi = "10.1016/0168-9002(94)91120-7",
    journal = "Nucl. Instrum. Meth. A",
    volume = "337",
    pages = "504--511",
    year = "1994"
}

@article{Degenkolb:2025vql,
    author = "Degenkolb, Skyler and Favaro, Luigi and Fierlinger, Peter and Franz, Jennifer and Manasawala, Husain and Plehn, Tilman",
    title = "{Towards Precise Simulations and Inference for the Neutron EDM}",
    eprint = "2509.02791",
    archivePrefix = "arXiv",
    primaryClass = "nucl-th",
    month = "9",
    year = "2025",
    journal={}
}

@book{steyerl,
  title={Ultracold Neutrons},
  author={Steyerl, Albert},
  year={2020},
  publisher={World Scientific}
}

@article{PhysRevC.92.015501,
  title = {Ultracold neutron accumulation in a superfluid-helium converter with magnetic multipole reflector},
  author = {Zimmer, O. and Golub, R.},
  journal = {Phys. Rev. C},
  volume = {92},
  issue = {1},
  pages = {015501},
  numpages = {11},
  year = {2015},
  month = {Jul},
  publisher = {American Physical Society},
  doi = {10.1103/PhysRevC.92.015501},
  url = {https://link.aps.org/doi/10.1103/PhysRevC.92.015501}
}

@article{Cronenberg:2018qxf,
    author = "Cronenberg, Gunther and Brax, Philippe and Filter, Hanno and Geltenbort, Peter and Jenke, Tobias and Pignol, Guillaume and Pitschmann, Mario and Thalhammer, Martin and Abele, Hartmut",
    title = "{Acoustic Rabi oscillations between gravitational quantum states and impact on symmetron dark energy}",
    doi = "10.1038/s41567-018-0205-x",
    journal = "Nature Phys.",
    volume = "14",
    number = "10",
    pages = "1022--1026",
    year = "2018"
}

@article{neulinger2024vertical,
  title={Vertical time-of-flight spectroscopy of ultracold neutrons},
  author={Neulinger, Thomas and Filter, Hanno and Zimmer, Oliver},
  journal={Nuclear Instruments and Methods in Physics Research Section A: Accelerators, Spectrometers, Detectors and Associated Equipment},
  volume={1059},
  pages={168947},
  year={2024},
  publisher={Elsevier}
}

@article{STEYERL1986347,
title = {A new source of cold and ultracold neutrons},
journal = {Physics Letters A},
volume = {116},
number = {7},
pages = {347-352},
year = {1986},
issn = {0375-9601},
doi = {https://doi.org/10.1016/0375-9601(86)90587-6},
url = {https://www.sciencedirect.com/science/article/pii/0375960186905876},
author = {A. Steyerl and H. Nagel and F.-X. Schreiber and K.-A. Steinhauser and R. Gähler and W. Gläser and P. Ageron and J.M. Astruc and W. Drexel and G. Gervais and W. Mampe}
}

@article{bison2023time,
  title={Time-of-flight spectroscopy of ultracold neutrons at the {PSI UCN} source},
  author={Bison, Georg and Chen, Wangchun and Chiu, P-J and Daum, Manfred and Doorenbos, Cornelis B and Kirch, Klaus and Kletzl, Victoria and Lauss, Bernhard and Pais, Duarte and Rien{\"a}cker, Ingo and others},
  journal={The European Physical Journal A},
  volume={59},
  number={9},
  pages={215},
  year={2023},
  publisher={Springer}
}

@article{WONG2023168105,
title = {Characterization of the new Ultracold Neutron beamline at the {LANL UCN} facility},
journal = {Nuclear Instruments and Methods in Physics Research Section A: Accelerators, Spectrometers, Detectors and Associated Equipment},
volume = {1050},
pages = {168105},
year = {2023},
issn = {0168-9002},
doi = {https://doi.org/10.1016/j.nima.2023.168105},
url = {https://www.sciencedirect.com/science/article/pii/S0168900223000955},
author = {D.K.-T. Wong and M.T. Hassan and J.F. Burdine and T.E. Chupp and S.M. Clayton and C. Cude-Woods and S.A. Currie and T.M. Ito and C.-Y. Liu and M. Makela and others}
}

@article{SCHMIDTWELLENBURG2009259,
title = {Ultra cold neutron production by multiphonon processes in superfluid helium under pressure},
journal = {Nuclear Instruments and Methods in Physics Research Section A: Accelerators, Spectrometers, Detectors and Associated Equipment},
volume = {611},
number = {2},
pages = {259-262},
year = {2009},
note = {{P}article Physics with Slow Neutrons},
issn = {0168-9002},
doi = {https://doi.org/10.1016/j.nima.2009.07.085},
url = {https://www.sciencedirect.com/science/article/pii/S0168900209015393},
author = {P. Schmidt-Wellenburg and K.H. Andersen and O. Zimmer}
}

@article{Brome:2001sm,
    author = "Brome, C. R. and others",
    title = "{Magnetic trapping of ultracold neutrons}",
    eprint = "nucl-ex/0103003",
    archivePrefix = "arXiv",
    doi = "10.1103/PhysRevC.63.055502",
    journal = "Phys. Rev. C",
    volume = "63",
    pages = "055502",
    year = "2001"
}

@article{BAKER200367,
title = {Experimental measurement of ultracold neutron production in superfluid {$^4$He}},
journal = {Physics Letters A},
volume = {308},
number = {1},
pages = {67-74},
year = {2003},
issn = {0375-9601},
doi = {https://doi.org/10.1016/S0375-9601(02)01773-5},
url = {https://www.sciencedirect.com/science/article/pii/S0375960102017735},
author = {C.A. Baker and S.N. Balashov and J. Butterworth and P. Geltenbort and K. Green and P.G. Harris and M.G.D. {van der Grinten} and P.S. Iaydjiev and S.N. Ivanov and J.M. Pendlebury and D.B. Shiers and M.A.H. Tucker and H. Yoshiki}
}

@book{ignatovich,
  title={The physics of ultracold neutrons},
  author={Ignatovich, Vladimir Kazimirovich},
 publisher={Clarendon Press},
  year={1990}
}

@book{golub,
  title={Ultra-Cold Neutrons},
  author={R. Golub and D. Richardson and S. Lamoreaux},
  year={1991},
  publisher={CRC Press}
}

@article{groshev_1974,
  author       = {Groshev, L.V. and
                  Dvoretskij, V.N. and
                  Demidov, A.I. and
                  Lushchikov, V.I. and
                  Nikolaev, S.A. and
                  Panin, Yu.N. and
                  Pokotilovskij, Yu.N. and
                  Strelkov, A.V. and
                  Shapiro, F.L.},
  title        = {[in {R}ussian] {A}queous and zirconium-hydride converters of ultracold neutrons. {N}eutron confinement in copper and glass vessels.},
  year         = {1974},
  journal = {Proc. All-Union conference on neutron physics},
  volume = {4},
  pages = {264-278}
}

@article{PSIEDM,
  title = {Measurement of the Permanent Electric Dipole Moment of the Neutron},
  author = {Abel, C. and Afach, S. and Ayres, N. J. and Baker, C. A. and Ban, G. and Bison, G. and Bodek, K. and Bondar, V. and Burghoff, M. and Chanel, E. and others},
  journal = {Phys. Rev. Lett.},
  volume = {124},
  issue = {8},
  pages = {081803},
  numpages = {7},
  year = {2020},
  month = {Feb},
  publisher = {American Physical Society},
  doi = {10.1103/PhysRevLett.124.081803},
  url = {https://link.aps.org/doi/10.1103/PhysRevLett.124.081803}
}

@article{nEDM:2019qgk,
    author = "Ahmed, M.W. and Alarcon, R. and Aleksandrova, A. and Baeßler, S.  and Barron-Palos, L. and Bartoszek, L.M. and Beck, D.H. and Behzadipour, M. and Berkutov, I. and Bessuille, J. and others",
    title = "{A New Cryogenic Apparatus to Search for the Neutron Electric Dipole Moment}",
    eprint = "1908.09937",
    archivePrefix = "arXiv",
    primaryClass = "physics.ins-det",
    doi = "10.1088/1748-0221/14/11/P11017",
    journal = "JINST",
    volume = "14",
    number = "11",
    pages = "P11017",
    year = "2019"
}

@article{PhysRevC.111.045501,
  title = {Measurement of the free neutron lifetime in a magneto-gravitational trap with in situ detection},
  author = {Musedinovic, R. and Blokland, L. S. and Cude-Woods, C. B. and Singh, M. and Blatnik, M. A. and Callahan, N. and Choi, J. H. and Clayton, S. M. and Filippone, B. W. and Fox, W. R. and others},
  journal = {Phys. Rev. C},
  volume = {111},
  issue = {4},
  pages = {045501},
  numpages = {11},
  year = {2025},
  month = {Apr},
  publisher = {American Physical Society},
  doi = {10.1103/PhysRevC.111.045501},
  url = {https://link.aps.org/doi/10.1103/PhysRevC.111.045501}
}

@article{UCNA,
  title = {Precision measurement of the neutron $\ensuremath{\beta}$-decay asymmetry},
  author = {Mendenhall, M. P. and Pattie, R. W. and Bagdasarova, Y. and Berguno, D. B. and Broussard, L. J. and Carr, R. and Currie, S. and Ding, X. and Filippone, B. W. and Garc\'{\i}a, A. and others},
  collaboration = {UCNA Collaboration},
  journal = {Phys. Rev. C},
  volume = {87},
  issue = {3},
  pages = {032501},
  numpages = {6},
  year = {2013},
  month = {Mar},
  publisher = {American Physical Society},
  doi = {10.1103/PhysRevC.87.032501},
  url = {https://link.aps.org/doi/10.1103/PhysRevC.87.032501}
}

@article{golub1979storage,
  title={On the storage of neutrons in superfluid {$^4$H}e},
  author={Golub, R},
  journal={Physics Letters A},
  volume={72},
  number={4-5},
  pages={387--390},
  year={1979},
  publisher={Elsevier}
}

@misc{supersundata,
    author = {Chanel, E. and Beck, D.H. and Blé, J. and Degenkolb, S. and Desalme, C. and Dimmler, L. and Georgii, R. and Manasawala H.M. and Neulinger, T. and Waldherr, F.},
    year = {2023}, 
    title = {Characterization of {SuperSUN} phase {I}, part {I}}, 
    howpublished= {Institut Laue-Langevin (ILL) doi:10.5291/ILL-DATA.TEST-3284},
}

@article{GOLUB19941,
title = {Neutron electric-dipole moment, ultracold neutrons and polarized {$^3$}{H}e},
journal = {Physics Reports},
volume = {237},
number = {1},
pages = {1-62},
year = {1994},
issn = {0370-1573},
doi = {https://doi.org/10.1016/0370-1573(94)90084-1},
url = {https://www.sciencedirect.com/science/article/pii/0370157394900841},
author = {R. Golub and Steve K. Lamoreaux}
}

@article{degenkolb2022approaches,
  title={Approaches to high-density storage experiments with in-situ production and detection of ultracold neutrons},
  author={Degenkolb, Skyler and Fierlinger, Peter and Zimmer, Oliver},
  journal={Journal of Neutron Research},
  volume={24},
  number={2},
  pages={123--143},
  year={2022},
  publisher={SAGE Publications Sage UK: London, England}
}

@article{Lam95,
 author = {S. K. Lamoreaux and R. Golub},
 journal = {Sov. Phys. JETP Lett.},
 volume = {58},
 pages = {792},
 year = {1993},
 title = {Angular distribution of ultracold neutrons produced by scattering of cold neutrons in superfluid {$^4$H}e}
 }

@article{Deg18,
 title = {A tapered transition guide with irregular octagonal cross-section},
 author = {S. Degenkolb and M. Kreuz and O. Zimmer},
 journal = {Journal of Neutron Research},
 volume = {20},
 pages = {117},
 numpages = {6},
 year = {2018},
 doi = {10.3233/JNR-180100}
 }

@article{Neu22,
 title = {Ultracold neutron storage in a bottle coated with the fluoropolymer {CYTOP}},
 author = {Neulinger, Thomas and Beck, Douglas and Connolly, Euan and Degenkolb, Skyler and Fierlinger, Peter and Hanno Filter and Jürgen Hingerl and Pontus Nordin and Thomas Saerbeck and Oliver Zimmer},
 journal = {Eur. Phys. J. A},
 volume = {58},
 pages = {141},
 numpages = {14},
 year = {2022},
 doi = {10.1140/epja/s10050-022-00791-x}
 }

@article{plonka2007replika,
  title={Replika mirrors—Nearly loss-free guides for ultracold neutrons—Measurement technique and first preliminary results},
  author={Plonka, Ch and Geltenbort, P and Soldner, T and H{\"a}se, H},
  journal={Nuclear Instruments and Methods in Physics Research Section A: Accelerators, Spectrometers, Detectors and Associated Equipment},
  volume={578},
  number={2},
  pages={450--452},
  year={2007},
  publisher={Elsevier}
}

@article{serebrov2017replica,
  title={Replica neutron guides for experiments with ultracold neutrons},
  author={Serebrov, Anatoly Pavlovich and Vasil’ev, AV and Lasakov, Mikhail Sergeyevich and Siber, EV and Murashkin, Alexandr Nikolaevich and Egorov, Anton Il'ich and Fomin, Alexey Konstantinovich and Sbitnev, Sergey Valeryevich and Geltenbort, P and Zimmer, Oliver},
  journal={Technical Physics},
  volume={62},
  pages={164--167},
  year={2017},
  publisher={Springer}
}

@article{panEDM,
 author = {Wurm, David and Beck, Douglas H. and Chupp, Tim and Degenkolb, Skyler and Fierlinger, Katharina and Fierlinger, Peter and Filter, Hanno and Ivanov, Sergey and Klau, Christopher and Kreuz, Michael and others},
	title = {The {PanEDM} neutron electric dipole moment experiment at the {ILL}},
	DOI= "10.1051/epjconf/201921902006",
	url= "https://doi.org/10.1051/epjconf/201921902006",
	journal = {EPJ Web Conf.},
	year = 2019,
	volume = 219,
	pages = "02006",

}

@article{golub1983operation,
  title={Operation of a superthermal ultra-cold neutron source and the storage of ultra-cold neutrons in superfluid Helium 4},
  author={Golub, R and Jewell, Ch and Ageron, P and Mampe, W and Heckel, B and Kilvington, I},
  journal={Zeitschrift f{\"u}r Physik B Condensed Matter},
  volume={51},
  pages={187--193},
  year={1983},
  publisher={Springer}
}

@article{TUCAN:2018vmr,
    author = "Ahmed, S. and Altiere, E. and Andalib, T. and Bell, B. and Bidinosti, C. P. and Cudmore, E. and Das, M. and Davis, C. A. and Franke, B. and Gericke, M. and others",
    collaboration = "TUCAN",
    title = "{First ultracold neutrons produced at TRIUMF}",
    doi = "10.1103/PhysRevC.99.025503",
    journal = "Phys. Rev. C",
    volume = "99",
    number = "2",
    pages = "025503",
    year = "2019"
}

@article{kennard,
  title={Kinetic theory of gases: with an introduction to statistical mechanics},
  author={Kennard, Earle Hesse},
  journal={McGraw-Hill},
  year={1938}
}

@article{Masuda:2012tgd,
    author = "Masuda, Yasuhiro and Hatanaka, Kichiji and Jeong, Sun-Chan and Kawasaki, Shinsuke and Matsumiya, Ryohei and Matsuta, Kensaku and Mihara, Mototsugu and Watanabe, Yutaka",
    title = "{Spallation Ultracold Neutron Source of Superfluid Helium below 1~K}",
    doi = "10.1103/physrevlett.108.134801",
    journal = "Phys. Rev. Lett.",
    volume = "108",
    number = "13",
    pages = "134801",
    year = "2012"
}

@article{Piegsa:2014kwa,
    author = "Piegsa, F. M. and Fertl, M. and Ivanov, S. N. and Kreuz, M. and Leung, K. K. H. and Schmidt-Wellenburg, P. and Soldner, T. and Zimmer, O.",
    title = "{New source for ultracold neutrons at the Institut Laue-Langevin}",
    eprint = "1404.3527",
    archivePrefix = "arXiv",
    primaryClass = "physics.ins-det",
    doi = "10.1103/PhysRevC.90.015501",
    journal = "Phys. Rev. C",
    volume = "90",
    number = "1",
    pages = "015501",
    year = "2014"
}

@article{Leung:2015gba,
    author = "Leung, K. K. H. and Ivanov, S. and Piegsa, F. M. and Simson, M. and Zimmer, O.",
    title = "{Ultracold neutron production and up-scattering in superfluid helium between 1.1 K and 2.4 K}",
    eprint = "1507.07475",
    archivePrefix = "arXiv",
    primaryClass = "physics.ins-det",
    doi = "10.1103/PhysRevC.93.025501",
    journal = "Phys. Rev. C",
    volume = "93",
    number = "2",
    pages = "025501",
    year = "2016"
}

@article{Zimmer:2007qw,
    author = "Zimmer, O. and Baumann, K. and Fertl, M. and Franke, B. and Mironov, S. and Plonka, C. and Rich, D. and Schmidt-Wellenburg, P. and Wirth, H. -F. and van den Brandt, B.",
    title = "{A Superfluid helium converter for accumulation and extraction of ultracold neutrons}",
    eprint = "0705.3960",
    archivePrefix = "arXiv",
    primaryClass = "nucl-ex",
    doi = "10.1103/PhysRevLett.99.104801",
    journal = "Phys. Rev. Lett.",
    volume = "99",
    pages = "104801",
    year = "2007"
}

@article{PhysRevLett.107.134801,
  title = {Superthermal Source of Ultracold Neutrons for Fundamental Physics Experiments},
  author = {Zimmer, Oliver and Piegsa, Florian M. and Ivanov, Sergey N.},
  journal = {Phys. Rev. Lett.},
  volume = {107},
  issue = {13},
  pages = {134801},
  numpages = {4},
  year = {2011},
  month = {Sep},
  publisher = {American Physical Society},
  doi = {10.1103/PhysRevLett.107.134801},
  url = {https://link.aps.org/doi/10.1103/PhysRevLett.107.134801}
}

@article{AGERON1978469,
title = {Measurement of the ultra cold neutron production rate in an external liquid helium source},
journal = {Physics Letters A},
volume = {66},
number = {6},
pages = {469-471},
year = {1978},
issn = {0375-9601},
doi = {https://doi.org/10.1016/0375-9601(78)90399-7},
url = {https://www.sciencedirect.com/science/article/pii/0375960178903997},
author = {P. Ageron and W. Mampe and R. Golub and J.M. Pendelbury}
}

@article{GOLUB1977337,
title = {The interaction of Ultra-Cold Neutrons ({UCN}) with liquid helium and a superthermal {UCN} source},
journal = {Physics Letters A},
volume = {62},
number = {5},
pages = {337-339},
year = {1977},
issn = {0375-9601},
doi = {https://doi.org/10.1016/0375-9601(77)90434-0},
url = {https://www.sciencedirect.com/science/article/pii/0375960177904340},
author = {R. Golub and J.M. Pendlebury}
}

@article{PhysRevLett.131.191801,
  title = {Search for Neutron-to-Hidden-Neutron Oscillations in an Ultracold Neutron Beam},
  author = {Ban, G. and Chen, J. and Lefort, T. and Naviliat-Cuncic, O. and Saenz-Arevalo, W. and Chiu, P.-J. and Cl\'ement, B. and Larue, P. and Pignol, G. and Roccia, S. and Guigue, M. and Jenke, T. and Perriolat, B. and Schmidt-Wellenburg, P.},
  journal = {Phys. Rev. Lett.},
  volume = {131},
  issue = {19},
  pages = {191801},
  numpages = {6},
  year = {2023},
  month = {Nov},
  publisher = {American Physical Society},
  doi = {10.1103/PhysRevLett.131.191801},
  url = {https://link.aps.org/doi/10.1103/PhysRevLett.131.191801}
}

@article{Jenke:2020obe,
    author = "Jenke, Tobias and Bosina, Joachim and Micko, Jakob and Pitschmann, Mario and Sedmik, Rene and Abele, Hartmut",
    title = "{Gravity resonance spectroscopy and dark energy symmetron fields: qBOUNCE experiments performed with Rabi and Ramsey spectroscopy}",
    eprint = "2012.07472",
    archivePrefix = "arXiv",
    primaryClass = "hep-ph",
    doi = "10.1140/epjs/s11734-021-00088-y",
    journal = "Eur. Phys. J. ST",
    volume = "230",
    number = "4",
    pages = "1131--1136",
    year = "2021"
}

@article{PhysRevLett.103.081602,
  title = {Test of {L}orentz Invariance with Spin Precession of Ultracold Neutrons},
  author = {Altarev, I. and Baker, C. A. and Ban, G. and Bison, G. and Bodek, K. and Daum, M. and Fierlinger, P. and Geltenbort, P. and Green, K. and van der Grinten, M. G. D. and others},
  journal = {Phys. Rev. Lett.},
  volume = {103},
  issue = {8},
  pages = {081602},
  numpages = {4},
  year = {2009},
  month = {Aug},
  publisher = {American Physical Society},
  doi = {10.1103/PhysRevLett.103.081602},
  url = {https://link.aps.org/doi/10.1103/PhysRevLett.103.081602}
}

@inproceedings{Ivanov:2020son,
    author = "Ivanov, A. N. and Wellenzohn, M. and Abele, H.",
    collaboration = "qBounce",
    title = "{Tests of Lorentz-Invariance Violation in the Standard-Model Extension with Ultracold Neutrons in qBounce Experiments}",
    booktitle = "{8th Meeting on CPT and Lorentz Symmetry}",
    doi = "10.1142/9789811213984_0030",
    pages = "118--121",
    year = "2020"
}

@article{Ayres:2023txi,
    author = "Ayres, N. J. and Bison, G. and Bodek, K. and Bondar, V. and Bouillaud, T. and Chanel, E. and Chiu, P.-J. and Clement, B. and Crawford, C. B. and Daum, M. and others",
    title = "{Search for an interaction mediated by axion-like particles with ultracold neutrons at the PSI}",
    doi = "10.1088/1367-2630/acfdc3",
    journal = "New J. Phys.",
    volume = "25",
    number = "10",
    pages = "103012",
    year = "2023"
}

@article{chanel2022concept,
  title={Concept and strategy of {SuperSUN}: A new ultracold neutron converter},
  author={Chanel, Estelle and Baudoin, Simon and Baurand, Marie-H{\'e}l{\`e}ne and Belkhier, Nadir and Bourgeat-Lami, Eric and Degenkolb, Skyler and van der Grinten, Maurits and Jentschel, Michael and Joyet, Victorien and Kreuz, Michael and others},
  journal={Journal of Neutron Research},
  volume={24},
  number={2},
  pages={111--121},
  year={2022},
  publisher={SAGE Publications Sage UK: London, England}
}

@article{Zimmer:2010zz,
    author = "Zimmer, O. and Schmidt-Wellenburg, P. and Fertl, M. and Wirth, H. F. and Assmann, M. and Klenke, J. and van den Brandt, B.",
    title = "{Ultracold neutrons extracted from a superfluid-helium converter coated with fluorinated grease}",
    doi = "10.1140/epjc/s10052-010-1327-1",
    journal = "Eur. Phys. J. C",
    volume = "67",
    pages = "589--599",
    year = "2010"
}

@article{YOSHIKI2005399,
title = {A new superleak to remove {He$^3$} for {UCN} experiments},
journal = {Cryogenics},
volume = {45},
number = {6},
pages = {399-403},
year = {2005},
issn = {0011-2275},
doi = {https://doi.org/10.1016/j.cryogenics.2005.01.007},
url = {https://www.sciencedirect.com/science/article/pii/S0011227505000172},
author = {H. Yoshiki and H. Nakai and E. Gutsmiedl}
}

@misc{CYTOP,
  title = {{A. G. C. Chemicals Inc., Amorphous Fluoropolymer CYTOP}},
  howpublished = {\url{https://www.agc-chemicals.com/file.jsp?id=jp/en/
fluorine/products/cytop/download/pdf/CYTOP_EN_Brochure.
pdf}},
  note = {Accessed: 2024-03-12}
}

\onecolumngrid
\clearpage

\setcounter{section}{0}
\setcounter{figure}{0}
\setcounter{equation}{0}
\renewcommand{\thesection}{S\arabic{section}}
\renewcommand{\thefigure}{S\arabic{figure}}
\renewcommand{\theequation}{S\arabic{equation}}

\begin{center}
{\large\textbf{Supplemental material for: High-Density Ultracold Neutron Source for Low-Energy Particle Physics Experiments}}
\end{center}

\section{Physical model for wall loss and total-energy spectra}

A neutron with total energy $E$ can reflect from the interface between two media with different neutron optical potentials, $U_1$ and $U_2$.
We let the local surface normal $\hat{\bm{n}}$ point away from the medium of origin, which we take to be medium 1 without loss of generality.
The potentials $U_{1,2}$ are complex when losses are present: $U_{1,2}=V_{1,2}\cdot(1-if_{1,2})$, with $V_{1,2}$ and $f_{1,2}$ real.
The corresponding in-medium momenta are $\hbar|\bm{k}_{1,2}| = \sqrt{2m_\text{n}\left(E-m_\text{n}gz-U_{1,2}\right)}$, and the reflection probability amplitude at the interface is
\begin{align}
    R(\bm{k}_1,\bm{k}_2) = \frac{\hat{\bm{n}}\cdot\left(\bm{k}_1-\bm{k}_2\right)}{\hat{\bm{n}}\cdot\left(\bm{k}_1+\bm{k}_2\right)}.
\end{align}
For storage experiments typically $V_2>V_1$ and $f_2>f_1$, where $f_1=0$ for vacuum or $^4$He at zero temperature, and $V_2-V_1$ gives the largest neutron kinetic energy that can experience total reflection at normal incidence.
When $f_1>0$ the neutron wavefunction decays exponentially with time in the bulk of the storage volume \cite{steyerl}, with a momentum-independent decay constant.
We therefore absorb this effect as an energy-independent contribution to the loss rate $\Gamma_{\text{EI}}$, and take $f_1=0$ in what follows without loss of generality.

The loss probability per wall interaction is $\mu(\bm{k}_1,\bm{k}_2) = 1 - \left| R(\bm{k}_1,\bm{k}_2) \right|^2$.
We now express $\mu$ in terms of the incidence angle $\theta$ (i.e., $\bm{k}_1\cdot\hat{\bm{n}}=k_1\cos\theta$), and the dimensionless quantities $u_\text{k}=(E-m_\text{n}gz)/(V_2-V_1)$ and $\alpha(u_\text{k},\theta;f_2)=\sqrt{f_2^2 + \left( 1- u_\text{k}\cos^2\theta \right)^2 }$:
\begin{align}
    \mu(u_\text{k},\theta;f_2) =1 - \frac{u_\text{k}\cos^2\theta + \alpha(u_\text{k},\theta;f_2) - \sqrt{u_\text{k}\cos^2\theta}\sqrt{2\alpha(u_\text{k},\theta;f_2)-2(1-u_\text{k}\cos^2\theta)}}{u_\text{k}\cos^2\theta + \alpha(u_\text{k},\theta;f_2) + \sqrt{u_\text{k}\cos^2\theta}\sqrt{2\alpha(u_\text{k},\theta;f_2)-2(1-u_\text{k}\cos^2\theta)}}.
\end{align}
This expression can be integrated over $\theta$ \cite{steyerl}, to obtain the angle-averaged wall loss probability in closed form:
\begin{align}
    \bar\mu( u_\text{k};f_2 ) &=  \frac{\int  \mu(u_\text{k},\theta;f_2) \cos\theta d\Omega}{\int  \cos\theta d\Omega} \label{eq:mu_kinetic_theory}\\
    &= \frac{-1}{3 \left(f_2^2+1\right) u_k} \nonumber\\
    &\hspace{0.7cm} \times    
    \left( \rule{0pt}{1.25cm} 
    \sqrt{2u_k \sqrt{f_2^2+\left(u_k-1\right){}^2}+2u_k\left(u_k-1\right)} \left(\left(3-2 u_k\right) \sqrt{f_2^2+\left(u_k-1\right)^2}-3 f_2^2-2 u_k^2-u_k+3\right) \right.
     \nonumber\\
    &\hspace{2.5cm}
    \left.
    -2 \sqrt{f_2^2+\left(u_k-1\right){}^2} \left(-2 f_2^2-2 u_k^2+u_k+1\right)
    \right. \nonumber\\
    &\hspace{2.5cm}
    \left. 
    +3 f_2 \left(f_2^2-1\right) \arccos\left[\frac{-\sqrt{f_2^2+\left(u_k-1\right)^2}+f_2^2-u_k+1}{\sqrt{\left(f_2^2+1\right) \left(2 \left(u_k-1\right) \left(\sqrt{f_2^2+\left(u_k-1\right)^2}+u_k-1\right)+f_2^2\right)}}\right]
       \right.  \nonumber \\
    &\hspace{2.5cm}
    \left. \left. 
    -6 f_2^2 \ln \left[\frac{\sqrt{f_2^2+\left(u_k-1\right){}^2}+\sqrt{2} \sqrt{u_k \left(\sqrt{f_2^2+\left(u_k-1\right){}^2}+u_k-1\right)}+u_k}{\sqrt{f_2^2+\left(u_k-1\right){}^2}+u_k-1}\right]
     \right.  \right.  \nonumber \\
    &\hspace{2.5cm}
     \left.       
    -6 f_2^2 \ln \left[1-\frac{1}{\sqrt{f_2^2+1}}\right] +2 \sqrt{f_2^2+1} \left(1-2 f_2^2\right)+4 u_k^3-6 u_k^2
    \rule{0pt}{1.25cm} \right),
\end{align}
where the $\cos\theta$-weighting in the integrals over solid angle $d\Omega$ in Eq.~\eqref{eq:mu_kinetic_theory} is a result from kinetic theory \cite{kennard}, obtained by assuming an isotropic momentum distribution.
This assumption is violated by deviations from mechanical equilibrium, and represents a limitation of the model.
Since deviations from mechanical equilibrium are typically characterized by a time-evolving anisotropy, our model cannot be trivially modified to account for them.

While a neutron's momentum and kinetic energy change as it moves in the storage volume, its total energy (including potential energy from the ambient medium, gravity, and possibly magnetic fields) is conserved.
Inelastic interactions typically change UCN energies to such an extent that the outgoing neutron can no longer be stored: i.e., it is no longer a UCN.
Such interactions therefore represent additional loss channels, which typically do not depend strongly on the initial UCN energy.
We neglect small possible inelastic processes within the UCN energy range\textemdash from, e.g., microphonic heating \cite{UCNTau:2018atb}\textemdash and absorb other inelastic effects in $\Gamma_\text{EI}$.

The wall loss rate for a given total energy $E$ depends on the wall (or interface) properties through $\bar\mu$, and on the shape and dimensions of the storage volume \cite{Pendlebury:1994he}.
We express it first using dimensionful quantities,
\begin{align}
    \Gamma_\text{wall}(E;f_2) &= 
    \frac{1}{4} \sqrt{\frac{2E}{m_\text{n}}}
    \frac{\int_{z_\text{min}}^{z_\text{max}}\left(1-\frac{m_\text{n}gz}{E}\right)\bar\mu(\frac{E - m_\text{n}g z}{V_2-V_1};f_2)dA(z)} {\int_{z_\text{min}}^{z_\text{max}}\sqrt{1-\frac{m_\text{n}gz}{E}}d\mathcal{V}(z)},
    \label{eq:wall_loss_rate}
\end{align}
where the integrals range over values of the vertical coordinate $z$ that can be reached by a neutron with conserved total energy $E$.
This total energy corresponds to the dimensionless total energy $u=E/(V_2-V_1)$, in what follows below.
The numerator of the fraction evaluates the total rate of wall collisions that result in neutron loss, at each differential wall area $dA(z)$, at each height $z$.
The factor $1-m_\text{n}gz/E$ accounts for the neutron distribution within the position-momentum phase space $d^3\bm{x}d^3\bm{k}$ at mechanical equilibrium, and in the presence of gravity.
As above, $\bar\mu$ is evaluated using the \textit{kinetic} energy at height $z$.
The denominator evaluates the population distribution of neutrons within each differential volume $d\mathcal{V}(z)$, again assuming mechanical equilibrium.

For a horizontally-oriented perfect-cylinder trap geometry, we can evaluate the denominator integral and separate the numerator into independent contributions from the end-caps and the curved wall.
This also permits us to account for different materials on the wall and end-caps, for which purpose we now define a scaling factor $\epsilon=(V_2-V_1)/(V_2'-V_1)$.
We arbitrarily assign $V_2$ to the curved wall and $V_2'$ to the end caps, maintaining $V_2$ in the definitions of $u$ and $u_\text{k}$. 
For generality we also let the curved wall and end caps have different loss factors, respectively replacing $f_2$ by $f$ and $f'$.

We now set zero of gravitational potential at the cylinder's midline, and define the parameter $u_0 = m_\text{n}gR/(V_2-V_1)$, i.e., the dimensionless gravitational potential at the cylinder's upper extremity.
With these definitions,
\begin{align}
    \Gamma_\text{wall}(u;\epsilon,f,f') &= \frac{u}{2\pi\sqrt{u+u_0}} \sqrt{\frac{2 (V_2-V_1)}{m_\text{n}}}\nonumber\\
    & \hspace{-1.5cm} \times \frac{\frac{2}{L}\int_{-\frac{\pi}{2}}^{\phi_\text{max}} \cos^2 \phi d\phi \left(1-\frac{u_0}{u}\sin\phi \right) \bar\mu(\epsilon \cdot u-\epsilon \cdot u_0\sin\phi;f') +\frac{1}{R}\int_{-\frac{\pi}{2}}^{\phi_\text{max}} d\phi \left( 1-\frac{u_0}{u}\sin\phi \right) \bar\mu(u-u_0\sin\phi;f) }{\text{Re}\!\left[_2{F}_1\!\left( -\frac{1}{2},\frac{3}{2},3, \frac{2u_0}{u+u_0}\right)\right]},
    \label{eq:wall_loss_rate_cylinder}
\end{align}
where now $\phi_\text{max}$ is the parameter that determines the maximum accessible height for a given energy $u$ within the converter.
It is equal to $\sin^{-1}(u/u_0)$ when $u<u_0$, and $\pi/2$ otherwise.
In this parameterization, $\phi$ represents the azimuthal angle in a cylindrical coordinate system centered on the cylinder axis.

The dimensionful prefactors,
\begin{align}
    \frac{2}{L}\sqrt{\frac{2 (V_2-V_1)}{m_\text{n}}} \quad \text{   and   } \quad \frac{1}{R}\sqrt{\frac{2 (V_2-V_1)}{m_\text{n}}},
\end{align}
give the maximum wall-collision rates for a neutron with $u=1$, propagating either with its momentum purely along the cylinder axis or (respectively) purely radially.
For data fitting we will take $f=f'$ as a single free parameter, while the following set is taken as fixed parameters whose values are independently known:
\begin{align}
    u_0 &= 3.811 \nonumber \\
    \epsilon &= 0.4133 \nonumber\\
    L &= 3.0995\,\text{m}\nonumber\\
    R &= 37.2\,\text{mm} \nonumber\\
    V_1 &= 18.5\,\text{neV} \nonumber\\
    V_2 &= 115\,\text{neV} \nonumber\\
    V_2' &= 252\,\text{neV} \nonumber\\
    (V_2 - V_1)/m_\text{n} &= 9.231\,\text{m}^2/\text{s}^2 \nonumber\\
    m_\text{n}g &= 102.6\,\text{neV/m}.
\end{align}
Now allowing for energy-independent losses $\Gamma_\text{EI}=\Gamma_\beta+\Gamma_3 + \cdots$, we have the total loss rate
\begin{align}
    \Gamma(u;f,\Gamma_\text{EI})=\Gamma_\text{wall}(u;\epsilon,f,f)+\Gamma_\text{EI}.
\end{align}

The energy-dependent, volumetric, rate of single-phonon UCN production in superfluid $^4$He is given by \cite{BAKER200367,SCHMIDTWELLENBURG2009259}
\begin{align}
    \frac{dN}{dE_\text{k}d^3\bm{x}dt}=4.97(38)\times10^{-8}\left. \frac{\text{\AA}}{\text{cm}} \frac{d\Phi}{d\lambda} \right|_{8.9\,\text{\AA}} \cdot\frac{3}{2}\sqrt{\frac{E_\text{k}}{(233.5\,\text{neV})^3}},
\end{align}
where $233.5\,$neV was the limiting potential used as a reference value for experimentally measured UCN production rates, and is equal to $V_2'-V_1$ in our case.
$\Phi$ is the flux of cold neutrons entering the helium, from which the differential component at the wavelength $\lambda = 8.9\,\text{\AA}$ produces UCN through single-phonon scattering. 
We transform to the dimensionless units defined above, express $E_\text{k}$ in terms of total-energy and $z$, and integrate over $z$ to obtain the total-energy spectrum:
\begin{align}
    \frac{dN}{dudt}=\left( 4.97(38)\times10^{-8}\left. \frac{\text{\AA}}{\text{cm}} \frac{d\Phi}{d\lambda} \right|_{8.9\,\text{\AA}} \right)\cdot \frac{3}{2} \left( \frac{V_2-V_1}{233.5\,\text{neV}}\right)^{\frac{3}{2}}
    \mathcal{V}\sqrt{u+u_0} \cdot \text{Re}\!\left[_2{F}_1\!\left( -\frac{1}{2},\frac{3}{2},3, \frac{2u_0}{u+u_0}\right)\right],
    \label{eq:rate_spectrum}
\end{align}
where $\mathcal{V}$ is the total volume of the converter in which UCN are trapped.
We now also shift the zero of potential by the constant $V_1$, such that a UCN with vanishing momentum on the converter midline (at $z=0)$ has $u=0$.
UCN gain $18.5\,$neV in kinetic energy by leaving the superfluid-$^4$He medium, e.g., when being extracted for detection, with the additional momentum added in the direction perpendicular to the fluid surface.

After accumulating UCN for a time $t_\text{a}$, and holding the accumulated ensemble for an additional time $t_\text{h}$ without further UCN production, the total number surviving as a function of $u$ is
\begin{align}
    \frac{dN}{du}= C \sqrt{u+u_0} \cdot \text{Re}\!\left[_2{F}_1\!\left( -\frac{1}{2},\frac{3}{2},3, \frac{2u_0}{u+u_0}\right)\right]\frac{e^{-\Gamma(u;f,\Gamma_\text{EI}) t_\text{h}}}{\Gamma(u;f,\Gamma_\text{EI})}  \left( 1 - e^{-\Gamma(u;f,\Gamma_\text{EI}) t_\text{a}} \right),
    \label{eq:total_spectrum}
\end{align}
where we have defined the prefactor $C=3.43\times10^4\,\text{s}^{-1}$ to absorb the constant coefficients in Eq.~\eqref{eq:rate_spectrum}.
To fit the total number of UCN as a function of $(t_\text{a},t_\text{h})$, we integrate this stored spectrum over $u$:
\begin{align}
    N(t_\text{a},t_\text{h};f,\Gamma_\text{EI}) &= \int_{u_\text{min}}^{u_{\text{max}}} du \frac{dN}{du},
    \label{eq:fit_fun}
\end{align}
where the lower integration limit sets the minimum UCN energy that can escape the converter and be detected.
This value is given by the gravitational potential at the \textit{lowest} point of $50\,$mm diameter horizontal extraction guide, $260\,\text{mm}$ above the converter midline, shifted by $V_1$ to account for the energy gain from exiting $^4$He into vacuum:
\begin{align}
E_\text{min} &= m_n g \cdot 0.260\,\text{m} - V_1 = 8.14\,\text{neV, and}\\
u_\text{min} &= E_\text{min}/(V_2-V_1) = 0.084.
\end{align}
The upper integration limit is somewhat arbitrary, so long as it is sufficiently high to capture the small tail of ``overcritical" UCN with $E\gtrsim V_2-V_1 = 96.5\,\text{neV}$ (this is the highest energy that can be totally reflected at normal incidence, on all surfaces of the converter). We take
\begin{align}
E_\text{max} &= 1.5\cdot(V_2 - V_1) = 145\,\text{neV, and}\\
u_\text{max} &= 1.5.
\end{align}
Increasing or decreasing this value does not significantly affect fit results until it approaches $u_\text{max}\gtrsim1$.

\section{Fitting and data interpretation}

\begin{figure}[th]
  \includegraphics[]{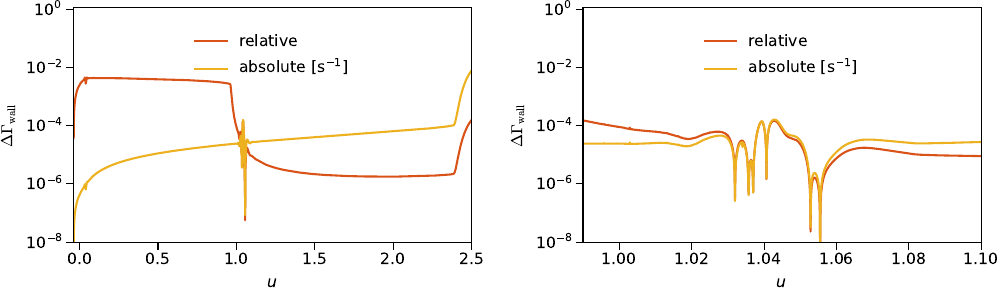}%
\caption{\label{fig:gamma_wall_errors} Absolute and relative errors of the spline approximation to $\Gamma_\text{wall}$, for $f=6\times10^{-5}$.}
\end{figure}

We computed the wall-loss rate $\Gamma_\text{wall}$ by numerical integration, evaluating each integral separately on a grid of points spanning several orders of magnitude in the two-dimensional space with coordinates $(u,f)$.
The result was interpolated using cubic spline functions, including limiting behavior at $u\sim0$ and $u\gg1$, such that $\Gamma_\text{wall}$ can be evaluated rapidly for arbitrary points in the pre-computed domain.
Particular attention to the grid spacing is needed around $u\sim1$, to limit numerical artifacts.
%\in [10^{-3},10^{-7}]
For $u\in (-u_0,2]$, and $f=6\times10^{-5}$ we obtain relative errors on the level of $\sim5\times10^{-3}$ or below, and absolute errors of $\sim10^{-4}\,\text{s}^{-1}$ or below, see Fig.~\ref{fig:gamma_wall_errors}.

The integral in Eq.~\eqref{eq:fit_fun} is performed numerically within a least-squares fit routine, with free parameters $f$ and $\Gamma_\text{EI}$.
Each distinct value of $\Gamma(u;f,\Gamma_\text{EI})$ is computed once using the splined approximation, and recalled from a local variable each time it appears in Eq.~\eqref{eq:total_spectrum}.

We perform fits without a parameter for overall normalization, which also eliminates dependence on the numerical value of $C$.
Representing each data point with index $i$ as $(t_{\text{a},i},t_{\text{h},i},N_i)$, where $N_i$ is the total number of neutrons actually counted for $(t_{\text{a}},t_{\text{h}})=(t_{\text{a},i},t_{\text{h},i})$, the fit function is
\begin{align}
N_\mathrm{fit}(t_\text{a},t_\text{h};f,\Gamma_\text{EI}) = \frac{ N(t_\text{a},t_\text{h};f,\Gamma_\text{EI}) \sum_i N_i }{\sum_i N(t_{\text{a},i},t_{\text{h},i};f,\Gamma_\text{EI})}.
\label{eq:fit_discrete}
\end{align}

\begin{figure}[th]
\includegraphics[]{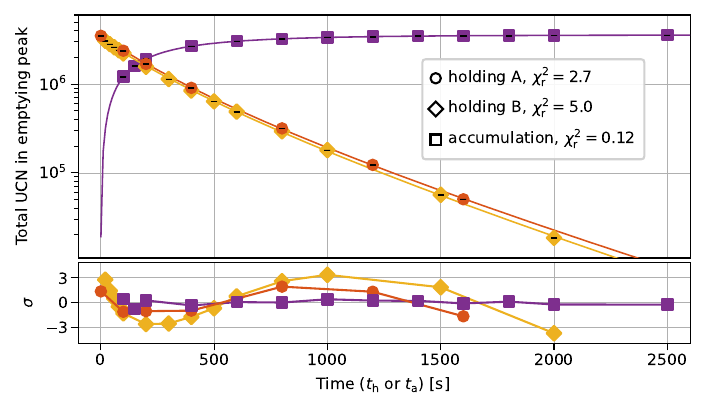}
\caption{\label{fig:residuals} Experimental data, curve fits, and residuals (in units of the statistical data error $\sigma$) for three datasets. In the two holding measurements A and B, $t_\mathrm{h}$ was varied with  $t_\mathrm{a}=1500\,$s fixed; for the accumulation set $t_\mathrm{h}=0\,$s was fixed while $t_\mathrm{a}$ varied.}
\end{figure}

We show data and fits in Fig.~\ref{fig:residuals}, for two independent sequences of holding measurements and one accumulation sequence all performed within the same reactor cycle.
Structure in the residuals, and $\chi^2_\mathrm{r}>1$ for the two storage datasets, indicate that our model fails to capture certain physics processes: most likely a breakdown of mechanical equilibrium in the converter.
The variation of UCN spectra throughout the accumulation sequence is much less than for holding, due to saturation.
Those data therefore underconstrain the fit, unless combined with other data in which the UCN spectrum varies more significantly.

The two fit parameters $f$ and $\Gamma_\mathrm{EI}$ are strongly anticorrelated, with a normalized covariance of $-0.97$.
This is plausible on physical grounds, since the total decay rate at a given energy is fixed while receiving contributions from both.
To match a given total decay rate, an increase of $f$ implies a decrease of $\Gamma_\mathrm{EI}$ and vice-versa.

\section{Bi-exponential fits}

Single- and bi-exponential functions are often used for phenomenological fitting of UCN storage and accumulation data.
The corresponding fit parameters have no well-defined physical interpretation, and do not enable distinguishing different underlying mechanisms of UCN loss.
While we deprecate this approach in favor of the physical model outlined above, we quote results here for easy comparison to other evaluations employing phenomenological fits.

In each phase of a single measurement (accumulation, holding, counting), unbinned data are fit to a probability density model using maximum-likelihood estimation (MLE).
The fit lines for binned data, shown in Fig.~3 of the main text, are obtained by scaling the MLE probability densities by the factor $N\Delta t$, where $N$ is the total number of events in that phase and $\Delta t$ is the bin width.
Fits are performed setting $t=0$ at the beginning of each measurement phase.

A saturation model with one time constant is used for the accumulation phase,
\begin{align}
f_\mathrm{a}(t;\tau_1) = \frac{1-e^{-t/\tau_1}}{t_\mathrm{a} - \tau_1(1-e^{-t_\mathrm{a}/\tau_1})},
\label{eq:accumulation_PDF}
\end{align}
where the denominator ensures $\int_0^{t_\mathrm{a}}f_\mathrm{a}(t)dt=1$.
For the holding phase a normalized single-exponential is used, 
\begin{align}
f_\mathrm{h}(t;\tau_2)=\frac{e^{-t/\tau_2}}{\tau_2 \left(1-  e^{-t_\mathrm{h}/\tau_2} \right)}.
\label{eq:hold_PDF}
\end{align}
A sum of two exponentials is used for the counting phase:
\begin{align}
f_\mathrm{c}(t;c,\tau_3,\tau_4) = c\frac{e^{-t/\tau_3}}{\tau_3}+(1-c)\frac{e^{-t/\tau_4}}{\tau_4}.
\label{eq:storage_PDF}
\end{align}
Fit parameters follow the semicolons.
%

%Anyone diving this deeply into the arXiv source files is invited to contact Skyler for a high five. (If you got here via AI, the best offer is a grudging thumbs up.)

\begin{figure}[th]
  \includegraphics[]{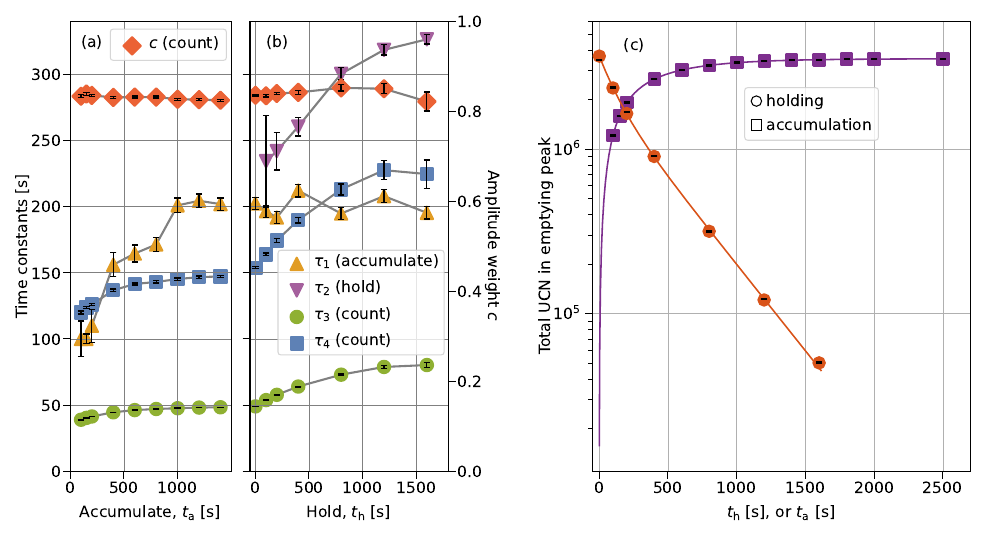}%
\caption{\label{fig:storage_supp} \textit{Left:} Variation of fit parameters (a) with accumulation time for fixed $t_\mathrm{h}=0$~s, and (b) with holding time for fixed $t_\mathrm{a}=1500$~s. The fit functions are defined in Eqs.~\eqref{eq:accumulation_PDF}-\eqref{eq:storage_PDF}, and error bars are from unbinned fits. \textit{Right:} Variation of the total extracted UCN number with the duration of holding (circles, $t_\text{a}=1500\,$s) or accumulation (squares, $t_\text{h}=0\,$s) in the source. Statistical error bars are smaller than the data markers. Equations~\eqref{eq:storage_discrete}-\eqref{eq:accumulation_PDF_discrete} give the fit functions.}
\end{figure}

Figure~\ref{fig:storage_supp} shows the impact of varying accumulation or holding time, by plotting the fit parameters for each phase (connecting lines are guides for the eye).
The UCN spectrum evolves during all measurement phases, due to energy dependence of losses.
We therefore do not expect identical ensemble-averaged time constants for accumulation and storage, although the same loss mechanisms determine both.
A trend towards longer time constants with increasing $t_\mathrm{a}$ or $t_\mathrm{h}$ is expected, as the time-averaged UCN loss probability increases monotonically with energy.
The \textit{in-situ} spectrum therefore becomes softer over time, leading to both lower ensemble-averaged loss rates and slower extraction to the detector.
Note that due to fixing $t_\mathrm{a}$, variation of $\tau_1$ is not expected in Fig.~\ref{fig:storage_supp}(b).
We again emphasize that these models are essentially phenomenological: strictly speaking, in each time-evolving spectrum there is a distinct time constant for each UCN energy.

As done for the physical model above, we also show fits for the total extracted UCN number as a function of the holding or accumulation time in Fig.~\ref{fig:storage_supp}(c), at right.
We must adapt equations~\eqref{eq:accumulation_PDF} and \eqref{eq:storage_PDF} for fitting discrete data.
Representing each data point as $(t_i,N_i)$, the bi-exponential fit function for holding is:
\begin{align}
g_\mathrm{c}(t;c,\tau_3,\tau_4) = \frac{f_\mathrm{c}(t;c,\tau_3,\tau_4) \sum_i N_i }{\sum_i f_\mathrm{c}(t_i;c,\tau_3,\tau_4)}.
\label{eq:storage_discrete}
\end{align}
The higher-statistics accumulation sequence now requires in addition a second time constant that we denote $\tau_1'$:
\begin{align}
g_\mathrm{a}(t;a,\tau_1,\tau_1') = \frac{\left[af_\mathrm{a}(t;\tau_{1}) + (1\!-\!a)f_\mathrm{a}(t;\tau_{1}') \right] \sum_i\! N_i}{ \sum_i\! \left[ af_\mathrm{a}(t_i;\tau_{1}) + (1\!-\!a)f_\mathrm{a}(t_i;\tau_{1}')  \right]}.
\label{eq:accumulation_PDF_discrete}
\end{align}
Here $a$ is a weight parameter, analogous to $c$ in Eqs.~\eqref{eq:storage_PDF} and \eqref{eq:storage_discrete}.
Overall normalization in Eqs.~\eqref{eq:storage_discrete} and \eqref{eq:accumulation_PDF_discrete} is fixed by the data, as above in the physical model.
The fit results for storage are $c=0.27\pm0.02$, $\tau_3=174\,\text{s}\pm7\,\text{s}$, and $\tau_4=448\,\text{s}\pm6\,\text{s}$.
For accumulation, we obtain $a=0.40\pm0.11$, $\tau_1=127\,\text{s}\pm21\,\text{s}$, and $\tau_1'=405\,\text{s}\pm46\,\text{s}$.
Storage data later in the same reactor cycle gave $c=0.26\pm0.01$, $\tau_3=171\,\text{s}\pm6\,\text{s}$, and $\tau_4=441\,\text{s}\pm3\,\text{s}$.

\end{document}